# Paired associative stimulation demonstrates alterations in motor cortical synaptic plasticity in patients with hepatic encephalopathy

Petyo **Nikolov**[1,2,*], Thomas J. **Baumgarten**[1,3,*], Shady S. **Hassan**[1,4], Sarah N. **Meissner**[1,5], Nur-Deniz **Füllenbach**[6], Gerald **Kircheis**[6], Dieter **Häussinger**[6], Markus S. **Jördens**[6], Markus **Butz**[1], Alfons **Schnitzler**[1,2], Stefan J. **Groiss**[1,2]

[1] Institute of Clinical Neuroscience and Medical Psychology, Medical Faculty, Heinrich Heine University, Düsseldorf, Germany

[2] Department of Neurology, Medical Faculty, Heinrich Heine University, Düsseldorf, Germany

[3] Neuroscience Institute, New York University School of Medicine, New York, USA

[4] Department of Neurology, Medical Faculty, Assiut University Hospital, Assiut, Egypt

[5] Department of Health Sciences and Technology, ETH Zürich, Switzerland

[6] Department of Gastroenterology, Hepatology and Infectious Diseases, Medical Faculty, Heinrich Heine University, Düsseldorf, Germany

*These authors contributed equally to this manuscript.

**Word count:** 6109; number of tables: 1; number of figures: 5; References: 94

**Short title:** PAS25 demonstrates altered plasticity over M1 in HE

**Address for Correspondence:**
Dr. Stefan Jun Groiss,
Department of Neurology, Heinrich Heine University Düsseldorf;
Moorenstraße 5; 40225 Düsseldorf, Germany.
email: groiss@uni-duesseldorf.de

[1]**List of abbreviations: AMPA**-Alpha-Amino-3-Hydroxy-5-Methyl-4-Isoxazole Propionic Acid, **APB**-abductor pollicis brevis, **CFF**-critical flicker frequency, **EMG**-electromyography, **GEE**-generalized estimating equations, **HE**-hepatic encephalopathy, **ICF**-intracortical facilitation, ISI-interstimulus interval, **mHE**-minimal hepatic encephalopathy, **HE1**-hepatic encephalopathy in grade one, **HE2**-hepatic encephalopathy in grade two, **LTD**-long-term depression, **LTP**-long-term potentiation, **M1**-motor cortex, **MEP**-motor evoked potential, **NMDA**-N-methyl-D-aspartate, **PAS**- paired associative stimulation, **PAS25**-paired associative stimulation with an inter-stimulus interval of 25 ms, **$PAS25_{LTP}$**- paired associative stimulation with an inter-stimulus interval of 25ms, resulting in long-term potentiation, **$PAS25_{LTD}$**- paired associative stimulation with an inter-stimulus interval of 25ms, resulting in long-term depression, **ROS**-reactive oxygen species, **RMT**-resting motor threshold, **STDP**-spike timing dependent plasticity, **SICI**: short-interval intracortical inhibition, TBS-theta-burst stimulation; cTBS-continuous theta-burst stimulation; iTBS-intermittent theta burst stimulation; **TMS**-transcranial magnetic stimulation

**Highlights:**

- Transcranial magnetic stimulation was used to study synaptic plasticity in patients with manifest hepatic encephalopathy.
- Patients with hepatic encephalopathy exhibited decreased synaptic plasticity of the primary motor cortex.
- This decrease may be caused by disturbances in the glutamatergic neurotransmission due to the known hyperammonemia.

**Abstract**

**Objective:** Hepatic encephalopathy (HE) is a potentially reversible brain dysfunction caused by liver failure. Altered synaptic plasticity is supposed to play a major role in the pathophysiology of HE. Here, we used paired associative stimulation with an inter-stimulus interval of 25 ms (PAS25), a transcranial magnetic stimulation (TMS) protocol, to test synaptic plasticity of the motor cortex in patients with manifest HE.

**Methods:** 23 HE-patients and 23 healthy controls were enrolled in the study. Motor evoked potential (MEP) amplitudes were assessed as measure for cortical excitability. Time courses of MEP amplitude changes after the PAS25 intervention were compared between both groups.

**Results:** MEP-amplitudes increased after PAS25 in the control group, indicating PAS25-induced synaptic plasticity in healthy controls, as expected. In contrast, MEP-amplitudes within the HE group did not change and were lower than in the control group, indicating no induction of plasticity.

**Conclusion:** Our study revealed reduced synaptic plasticity of the primary motor cortex in HE.

**Significance:** Reduced synaptic plasticity in HE provides a link between pathological changes on the molecular level and early clinical symptoms of the disease. This decrease may be caused by disturbances in the glutamatergic neurotransmission due to the known hyperammonemia in HE patients.

**Key words:** transcranial magnetic stimulation, paired associative stimulation, hepatic encephalopathy, critical flicker frequency, motor evoked potential, synaptic plasticity.

## 1. Introduction

Hepatic encephalopathy (HE) is a potentially reversible brain dysfunction due to liver failure (Häussinger et al., 2021; Häussinger and Schliess, 2008; Häussinger and Sies, 2013), which occurs in about 30 to 45 % of patients with liver cirrhosis (Vilstrup et al., 2014). HE comprises a broad spectrum of neuropsychiatric impairments, including motor impairments, deficits in visual and tactile perception, cognitive dysfunction, and impaired consciousness (Felipo, 2013; Henderson, 2008; Kircheis et al., 2002; Lazar et al., 2018). While historically, HE has been viewed as a single syndrome, the diversity of symptoms is currently thought to be mediated by different underlying mechanisms (Felipo, 2013). Yet, it is generally agreed that both systemic inflammation and hyperammonemia are crucial for symptom development (Coltart et al., 2013; Desjardins et al., 2012; Rose, 2012). Although potentially reversible, manifest HE often results in frequent and recurring hospitalization and thus constitutes a substantial burden for both patients and healthcare systems (Stepanova et al., 2012).

Previous work demonstrated decreased cortical synaptic plasticity in HE (Wen et al., 2013). Presumably, such alterations result from oxidative stress in neurons and glial cells due to accumulation of neurotoxins, especially ammonia (Häussinger and Schliess, 2008). Specifically, hyperammonemia has been connected to reductions in extrasynaptic reserve pools of AMPA-type glutamate receptors, which in turn severely limits synaptic plasticity (Schroeter et al., 2015). Importantly, alterations in synaptic plasticity are interpreted as precursor for the global cortical slowing of neuronal oscillatory activity (Butz et al., 2013; May et al., 2014; Timmermann et al., 2005), which is suggested to lead to the perceptual (Baumgarten et al., 2018; Götz et al., 2013) and motor deficits (Cantarero et al., 2013; Timmermann et al., 2008) present in HE. Thus, cortical synaptic plasticity changes are thought to represent a key mechanism connecting disease-related effects on the molecular level with impaired behavioral parameters. Although different forms of cortical synaptic plasticity have been described, the present work focusses on long-term potentiation (LTP; Malenka and Bear, 2004), since this process can be experimentally assessed *in vivo* in HE patients directly by means of non-invasive brain stimulation methods.

Transcranial magnetic stimulation (TMS) offers a non-invasive and painless method to investigate cortical physiology (Hallett, 2000; Rothwell, 1997). Paired associative stimulation (PAS; Stefan et al., 2002, 2000), an experimental TMS paradigm combining afferent electrical and cortical magnetic stimulation in a precise temporal regime, allows for the investigation of motor cortical plasticity in particular. To this end, an electrical stimulation of the median nerve is temporally paired with a TMS pulse over the contralateral primary motor cortex (M1). In one variant of PAS, median nerve stimulation is applied 25 ms before the TMS pulse, a protocol that is referred to as PAS25. As the afferent input from the median nerve needs 21 - 23 ms to reach M1, the electrically-induced neural signal arrives at M1 immediately before the TMS pulse (Wolters et al., 2003). Hence, the PAS25 protocol is known to enhance excitability within M1, where excitability evolves rapidly and remains present for an extended duration, inducing long-term potentiation (LTP) of synaptic efficacy ($PAS25_{LTP}$). Conceptually, an increase in excitability is interpreted as spike timing dependent plasticity (STDP), which is considered as one of the core mechanisms driving LTP (Stefan et al., 2002, 2000). Consistently, long-term depression (LTD) effects are also known to occur when the afferent input from the median nerve arrives at M1 after the TMS pulse (Di Lazzaro et al., 2009; Wolters et al., 2003).

While the PAS25 protocol offers an established and mechanistically well understood option to non-invasively study motor cortical plasticity in healthy subjects and patient populations (Golaszewski et al., 2016), motor cortical plasticity has been investigated with various different TMS protocols, e.g., PAS21.5 (Hamada et al., 2012) or intermittent theta-burst stimulation (Huang et al., 2005). Although both PAS21.5 and PAS25 induce LTP-like changes in the motor cortex, subsequent studies have highlighted different pathways underlying these effects (Hamada et al., 2012; Popa et al., 2013; Strigaro et al., 2014). Whereas PAS21.5 is thought to rely on signal transduction by means of a direct sensory pathway to the motor cortex, PAS25 is mediated by comparatively complex circuits (Hamada et al., 2012), including slow extra-lemniscal relays and cerebellar inputs (Butler et al., 1992; Popa et al., 2013). In the present study, we selected the PAS25 protocol as it represents the most frequently used PAS protocol variant to elicit LTP-like effects (reviewed in Wischnewski and Schutter, 2016). Furthermore, the complex circuit dynamics involved in PAS25 allow for investigation of a wide range of plasticity mechanisms potentially impaired in HE. Such a global approach is motivated by the broad symptom spectrum associated with HE.

Although PAS protocols offer an intriguing possibility to non-invasively study motor cortical plasticity in patient populations (Golaszewski et al., 2016), earlier reports describe considerable inter-individual variability in responses to PAS. Regarding PAS25 for example, about half of the subjects are known to not respond to the study protocol (López-Alonso et al., 2014). Because of this high variability, it is possible that up to 50% of the subjects even develop signs of LTD ($PAS25_{LTD}$), opposing general protocol expectations (Müller-Dahlhaus et al., 2008).

Here, we investigated synaptic plasticity of M1 in patients with manifest HE in different stages of disease severity and healthy controls. We hypothesized that healthy participants demonstrate LTP-like effects after PAS25 intervention, while such effects would be diminished or absent in HE patients.

## 2. Materials & Methods

### 2.1. Participants

23 patients (15 male, 8 female; age: 60.83 ± 1.35 (mean ± SEM)) with hepatic encephalopathy (HE) and 23 healthy volunteers (13 male, 10 female; age: 61.45 ± 1.46) participated in the present study (see **Tab. 1** for details). All participants provided written informed consent prior to study participation. The study was performed in accordance with the Declaration of Helsinki (World Medical Association, 2013) and was approved by the ethics committee of the Medical Faculty, Heinrich Heine University Düsseldorf, Germany (study number: 5179R). Patient inclusion criteria were clinically confirmed liver cirrhosis and the diagnosis of HE. Grading of HE severity was based on the *West-Haven* criteria (Ferenci et al., 2002), the critical flicker frequency (Kircheis et al., 2014, 2002), and a clinical assessment of the mental state and consciousness assessed by an experienced clinician. Healthy volunteers were recruited as an age-matched control group.

Exclusion criteria for both patients and controls included contraindications to TMS (e.g., due to metallic and/or magnetic implants), severe intestinal, neurological, or psychiatric diseases (except the diagnosis of HE for the patient group), the use of any medication acting on the central nervous system (e.g., benzodiazepines, anti-epileptic and/or psychotropic drugs), blood clotting dysfunction, pregnancy, and diagnosed peripheral/retinal neuropathy. Further, patients had to confirm alcohol abstinence for ≥ 4 weeks prior to measurement.

### 2.2. Experimental design

Participants were seated in a comfortable reclining chair with arms placed on cushioned armrests during the entire experiment. Recording of the individual critical flicker frequency was performed prior to the TMS-protocol. Electromyographic electrodes were attached to the right abductor pollicis brevis (APB) in belly-tendon montage. Then, the individual location for TMS (i.e., the "hotspot") was determined in steps of 0.5 to 1 cm (starting 5 cm lateral and 1.5 cm anterior from the vertex) at the stimulation site at which moderately supra-threshold TMS resulted in the largest consistent APB MEP amplitude. Here, the MEP-amplitude was defined as consistent, if it remained the same in at least 5 out of 10 trials. The hotspot was marked directly on the scalp to ensure constant placement of the TMS coil throughout the session. Based on this hotspot, TMS intensities for resting motor threshold (RMT), as well as for evoking MEP of 0.5 mV amplitude were determined. Next, the electrodes for the electrical stimulation of the median nerve were positioned on the right wrist and the individual perceptual threshold was determined.

MEP recordings were performed in four subsequent sessions: (i) Before paired associative stimulation (PAS) serving as baseline, (ii) 5 min after PAS, (iii) 15 min after PAS, and (iv) 25 min after PAS (**Fig. 1**). To ensure comparability, all measurements were carried out during the early afternoon (i.e., between 1:00 pm and 3:00 pm), since cortical plasticity is known to be influenced by circadian rhythms (Sale et al., 2007). Through the course of the entire experiment, muscle relaxation was visually monitored online on an oscilloscope. Participants were instructed to look at a fixation cross centered in front of them and silently count the number of magnetic pulses applied to maintain similar levels of attention.

### 2.3. Electromyographic recording (EMG)

EMG signals were recorded from the right APB muscle by means of disposable Ag-AgCl surface electrodes (20 x 15 mm, Ambu A/S, Denmark). The active electrode was placed on the muscle belly, whereas the reference was placed over the base of the metacarpophalangeal joint of the thumb. EMG signals were amplified (Digitimer D360, Digitimer Ltd, Hertfordshire, UK), band-passed between 100 Hz and 5 kHz, digitized at a sampling rate of 5 kHz, and stored on a desktop computer for later off-line analysis.

### 2.4. Electrical stimulation

Electrical stimulation consisted of a square wave pulse with 0.2 ms duration and was applied to the right median nerve by a Digitimer Constant Current Stimulator (DS7A, Digitimer Ltd, Hertfordshire, UK). The electrical stimulation was delivered with a pair of disposable Ag-AgCl surface electrodes (20 x 15 mm, Ambu A/S, Denmark). Electrodes were positioned on the palmar side of the wrist, proximal to the metacarpophalangeal joint. 300% the amount of the previously determined individual perceptual threshold was used for the electrical stimulation in the following PAS25 intervention. To ensure that electrical stimulation was above motor threshold, it was verified by visual conformation of muscle twitch (Stefan et al., 2000).

### 2.5. Transcranial magnetic stimulation (TMS)

TMS was applied by a Magstim™ magnetic stimulator (The Magstim Co. Ltd, Whitland, UK) in combination with a figure-of-eight coil. The coil was placed above the left primary motor cortex (M1) with hotspot APB tangentially to the scalp with the coil handle pointing backwards and laterally at a 45° angle to the sagittal plane in order to ensure a posterior-anterior current direction in the brain (Rothwell, 1997). This configuration aims to trans-synaptically activate the corticospinal system by means of horizontal cortico-cortical connections (Di Lazzaro et al., 2004). After determination of the individual TMS hotspot, the resting motor threshold (RMT) was defined as the lowest stimulation intensity that evoked a response of at least 50 µV during complete relaxation of the right APB in at least 5 of 10 trials using the relative frequency method (Rossini et al., 2015).

### 2.6. Critical flicker frequency (CFF)

CFF was measured with a mobile device (HEPAtonorm$^{TM}$-Analyzer, nevoLAB, Maierhöfen, Germany) just prior to the TMS measurement. Assessment of the individual CFF is based on the presentation of a flickering light, which starts to flicker with a frequency of 60 Hz. At 60 Hz, the light is perceived as a continuous stimulus. Subsequently, the frequency with which the light flickers decreases linearly. Participants are requested to report when they first perceive the light as clearly flickering. Participants were first trained in this process and after they confirmed to understand the instructions, the measurement was repeated three times and the mean CFF value was taken for further analysis. The individual CFF is known to decrease depending on HE disease severity, with 39 Hz suggested as a cut-off to detect minimal HE in patients (Kircheis et al., 2014, 2002).

### 2.7. Paired associative stimulation (PAS)

The PAS25 protocol applied was based on the protocol of Delvendahl et al. ( 2012). In accordance, we applied 180 low-frequency (0.2 Hz) single-pulse TMS of the left M1 paired with 180 electrical stimuli of the right median nerve (Suppa et al., 2017). Electrical and magnetic stimulation were separated by an interstimulus interval of 25 ms. Since MEP recruitment curve is known to be shallower in HE patients (Groiss et al., 2019), TMS intensity was adjusted to reliably evoke MEP with an amplitude of 0.5 mV.

### 2.8. Motor-evoked potential recordings

MEP responses evoked by single pulse TMS, amplitude-adjusted to evoke MEPs of 0.5 mV at baseline, were recorded at each of the four sessions and its amplitudes were used as a measure to indicate cortical excitability. In addition, recruitment curves were measured at each of the four sessions. Recruitment curves were generated by five different stimulation intensities, i.e., at 100%, 110%, 120%, 130%, and 140% of the individual RMT. The order of stimulation conditions was randomized across and within each participant. For each stimulation intensity, 12 MEPs were recorded during each of the four sessions. Within each session, MEPs were averaged across trials for each stimulation intensity. In total, 72 MEPs were recorded during each session.

EMG data was analyzed with Signal Software (Cambridge Electronic Design, Cambridge, UK). Trials were visually inspected for artifacts. Trials showing voluntary EMG activity immediately before the TMS pulse, as well as trials where TMS stimulation was missing due to technical reasons, were rejected from the analysis. Maximum peak-to-peak MEP amplitudes were determined for each trial. Subsequently, peak-to-peak amplitudes were averaged over all trials within a session.

### 2.9. Definition of $PAS25_{LTP}$ and $PAS25_{LTD}$

In general, the expected response to PAS25 is an MEP increase in sessions following PAS25 application, reflecting LTP-response (**$PAS25_{LTP}$**). However, about a quarter of tested individuals are known to rather respond with an MEP decrease (**$PAS25_{LTD}$**), while others show no MEP change at all (non-responders; Müller-Dahlhaus et al., 2008; Nakatani-Enomoto et al., 2012; Wiethoff et al., 2014). To allow a precise analysis of LTP-responses while considering the substantial inter-individual variability in PAS25 responses across individuals, we categorized participants depending on their respective MEP responses into LTP-responders, LTD-responders, or non-responders (**Fig. 2**). In a first step, participants were categorized as responders if their average MEP amplitude change across trials exceeded baseline by 0.2 mV in any one of the three sessions after PAS25 intervention. Notably, this cut-off value is defined by an increase larger than 2 SDs of the inter-trial variability found at baseline (Campana et al., 2019; Nakamura et al., 2016; Tiksnadi et al., 2020). If this criterion was not met, the respective participant was categorized as non-responder. Within the responder group, we separated participants into either $PAS25_{LTP}$ or $PAS25_{LTD}$ responders, depending on the direction of average amplitude change relative to baseline (i.e., increase or decrease) in those sessions that exceeded the baseline by ≥0.2 mV. In cases where participants exhibited both increases and decreases surpassing baseline by ≥0.2 mV across different sessions, categorization was based on the maximum absolute amplitude change.

### 2.10. Statistical evaluation

Statistical evaluation was performed with Graph pad Prism5™ (GraphPad Software, San Diego, USA) and SPSS (Version 25, IBM, Armonk, USA). Participant age was compared between patients and controls with a two-sample t-test. CFF and RMT were compared between patients and controls using a Mann-Whitney U test. To determine a potential influence of age or perceptual threshold, age and perceptual threshold were correlated with CFF and relative MEP amplitude change after PAS25-intervention. All correlations were carried out with the Spearman rank correlation test. The Shapiro-Wilk test was used to test the distributions of recorded parameters for normality, determining the use of either parametric or nonparametric statistical tests. Comparison of LTP-responder rate between controls and HE patients was carried out with the $Chi^2$ test of independence. For analyses of demographic parameters, all participants (i.e., $PAS25_{LTP}$-responders, $PAS25_{LTD}$-responders, and non-responders) were included in the analysis.

To determine the effect of PAS25 on LTP in the patient and control group, respectively, we compared MEP amplitude changes relative to baseline between the four sessions (baseline, 5 min, 15 min, 25 min) using the Friedman test. If applicable, we used Dunn test for post-hoc analysis. Here, only $PAS25_{LTP}$-responders were included in the analysis.

To compare relative MEP amplitude changes after PAS25 intervention between the patient and control group, we first calculated MEP ratios at 5 min-, 15 min-, or 25 min-post PAS25 relative to baseline for both controls and HE-patients, respectively. Thus, MEP ratios represent the degree of $PAS25_{LTP}$ response over time. Next, we compared the degree of $PAS25_{LTP}$ response between controls and HE-patients at the 5 min post PAS session using a two-sample t-test. Here, only the $PAS25_{LTP}$-responders were included in the analysis.

Next, we determined if the presence of HE has an influence on the magnitude of MEP amplitude change due to PAS25 intervention. To this end, we calculated the absolute MEP ratio difference from baseline (i.e., |MEP ratio per post-PAS25-session - 1|), independently of the direction on amplitude change. Thus, the output quantifies the overall magnitude of MEP amplitude change due to PAS25 intervention, while not differentiating LTP- from LTD-responses. The $PAS25_{LTP}$ or $PAS25_{LTD}$ response was analyzed with linear generalized estimating equations (GEE). Here, the absolute MEP ratio difference from baseline was set as dependent variable, while presence of HE and session were set as co-factors. Both $PAS25_{LTP}$ and $PAS25_{LTD}$-responders were included in the analysis.

Finally, we compared recruitment curves across participants. Again, a linear GEE-model was used, with MEP-amplitude as dependent variable. Session, intensity, and presence of HE were included as co-factors. All participants (i.e., $PAS25_{LTP}$-responders, $PAS25_{LTD}$-responders, and non-responders) were included in the analysis.

## 3. Results

### 3.1. Demographic Data

Seven of the 23 HE-patients enrolled in the study exhibited minimal hepatic encephalopathy (mHE); 13 were in grade one (HE1), and three in grade two (HE2) according to *West-Haven* criteria. All patients in the study were diagnosed with a liver cirrhosis for at least two months prior inclusion. There was no significant difference in age, as well as perceptual threshold, and RMT between HE-patients and healthy controls (p>0.05 for all comparisons). Age and perceptual threshold of both patients and controls also did not correlate significantly with either RMT or MEP-change after PAS25-intervention (*p*>0.05 for both correlations). On average, we discarded 10 MEP trials per participant (from a total of 288 trials across all sessions per participant) due to signal artifacts.

### 3.2. Critical flicker frequency

CFF could be successfully determined in all HE-patients and in 11 healthy controls. CFF was significantly lower in HE-patients (36.72±0.54 Hz (mean ± SEM)) than in the control group (42.06±0.58 Hz; U = 9, *p*<0.001).

### 3.3. $PAS25_{LTP}$ response

$PAS25_{LTP}$-responder rate did not differ significantly between HE patients and healthy controls (*p*>0.05; see **Tab. 1** for further details). In the control group, MEP amplitudes were significantly different between sessions ($F_R$=13.8, *p*<0.01). *Post-hoc* analysis revealed MEP amplitudes pre-PAS25 (at baseline, mean MEP=0.53±0.15 mV) to be significantly smaller than 5 min post PAS25 (mean MEP=2.34±0.26 mV, *p*<0.01), while the 15 min and the 25 min post PAS sessions did not significantly differ from baseline or among each other (all *p*>0.05). In the HE group, there was no significant difference of MEP amplitudes between sessions (all *p*>0.05, **Fig. 3**).

MEP ratio at 5 min post PAS in the control group (mean MEP=2.34±0.26 mV) was significantly higher than in the HE group (mean MEP=1.13±0.14 mV, *p*<0.01, **Fig. 3**, inset).

### 3.4. $PAS25_{LTP}$ or $PAS25_{LTD}$ response

Regarding $PAS25_{LTP}$ or $PAS25_{LTD}$ response, we found the presence of HE to be a significant factor ($p$=0.001), determining plasticity change in both directions (mean absolute plasticity change for controls=0.64±0.07 mV; for HE-patients=0.37±0.04 mV). Session, on the other hand, was not a significant factor ($p$>0.05).

### 3.5. Recruitment curves

Regarding the recruitment curves, we found the presence of HE to be a significant factor determining MEP amplitude ($p$<0.01). Mean MEP-amplitudes during recruitment curves in controls were 0.78±0.04 mV, while amplitudes were 0.41±0.03 mV in HE patients. Taken together, MEPs in recruitment curves of HE patients were significantly smaller, as compared to healthy controls (**Fig. 4**).

As expected, TMS-intensity was a significant factor in the recruitment curves analysis ($p$<0.001, mean MEP at 100%RMT=0.11±0.01 mV; at 110%RMT= 0.25±0.02 mV; at 120%RMT=0.56±0.05 mV; at 130%RMT=0.89±0.06 mV; at 140%RMT=1.17±0.07 mV). Session was not a significant factor for MEP amplitudes during recruitment curves measurement ($p$=0.52).

## 4. Discussion

Our study has two main findings. First, significant PAS-induced long term-potentiation (LTP) in M1 could not be observed in HE patients, but exclusively in healthy controls. In addition, MEP change post PAS25 was significantly higher in healthy controls than in HE patients. Second, MEP amplitudes of recruitment curves in HE patients were significantly smaller, as compared to those of healthy controls, which indeed confirmed our previous findings. Taken together, we could demonstrate reduced cortical plasticity using PAS25 protocol in patients suffering from HE, including patients with manifest HE, for the first time.

The reduced plasticity described here tallies with earlier findings on cortical plasticity in patients with mHE. Golaszewski et al. (2016) applied PAS25 protocol on 15 patients with minimal HE. While a significant increase in MEP amplitude after PAS25 occurred in the control group, no significant change in MEP amplitude after PAS25 was observed in the mHE group. While these results are in line with the present findings and generally support a differential effect of PAS25 on HE patients and participants not affected by HE, some important methodological differences have to be highlighted. First, Golaszewski et al. (2016) report results for patients with minimal (i.e., subclinical) HE and cirrhotic patients without HE serving as control group. In contrast, our patient sample included mHE, HE1 (i.e., overt HE), and HE2 patients, whereas controls were recruited from healthy (i.e., non-cirrhotic) volunteers. Thus, while the results of Golaszewski et al. (2016) and others (Nardone et al., 2016) indicate that impaired plasticity is a specific hallmark of HE and not generally associated to cirrhosis, our results demonstrate that reduced plasticity is present across HE disease stages. On the other hand, Golaszewski et al. (2016) did not differentiate their participants based on their individual direction of MEP amplitude change after PAS25 stimulation. This potential mixing of LTP- and LTD- responders makes it difficult to determine if group-level results indicating the absence of MEP-changes after PAS stimulation are caused by a general lack of MEP-response (i.e., an increase in non-responders) or an altered ratio of LTP- and LTD-responders. In the present study, we categorized participants based on their post PAS25 MEP response, thereby focusing exclusively on LTP-responders when assessing post PAS25 MEP amplitude changes. Thus, even when restricting analysis exclusively to LTP-responders, our study demonstrates impaired synaptic plasticity in HE, indicating that plasticity changes in HE cannot solely be attributed to changes in the ratio of LTP- vs. LTD-responders.

Importantly, the distinction between responders and non-responders, as well as between LTP- and LTD-like responses requires the use of a categorization threshold. While earlier studies mostly did not distinguish between responders and non-responders and determined responses to be LTP- vs. LTD-like based solely on the how MEP amplitudes changed compared to baseline (i.e., increase or decrease; e.g., Müller-Dahlhaus et al., 2008), we required participants to show a mean MEP amplitude change of ≥0.2 mV before categorizing them as responders. This method was motivated by recent studies (e.g., Campana et al., 2019; Nakamura et al., 2016; Tiksnadi et al., 2020) and aims to provide a more robust assessment of TMS-induced responses. Specifically, the rationale is that responses are only taken into account if they surpass 2 standard deviations of baseline inter-trial MEP variability. It is possible that statistical comparison between MEPs recorded in baseline vs. post PAS-sessions might yield an even more robust categorization threshold and this should be validated as further categorisation method in future TMS studies.

The reduced synaptic plasticity in HE, as assessed by TMS, might serve as an important hint at the molecular mechanisms behind this disease. It is largely accepted that synaptic plasticity and LTP are mainly generated by repetitive activation of glutamatergic synapses (Classen et al., 2004; Mansvelder et al., 2019). Therefore, impairment of the glutamatergic neurotransmission might well explain the reduced LTP in HE. However, the question remains how glutamatergic neurotransmission is specifically impaired in HE. Here, hyperammonemia seems to play a major role. In the human metabolism, ammonium ions are constantly produced and consumed. A main source of ammonium is the gastro-intestinal tract, where it is either produced by bacteria, or extracted during diet uptake in the form of glutamine. A proper ammonium homeostasis is crucial for multiple body functions and the liver is the key organ involved in its maintenance. In a cirrhotic liver, however, ammonium clearance is impaired, causing hyperammonemia, an excess of ammonia in the blood. In such conditions, cerebral ammonium uptake increases linearly with its concentration in arterial blood (Olde Damink et al., 2002).

Cerebral hyperammonemia is closely linked to glutamate metabolism and neurotransmission. For example, one of the major ammonium detoxification mechanisms of the brain involves glutamine synthesis by binding superfluous ammonium to glutamate. Hereby, the glutamine synthesis takes place in the astrocytes' cytosol (Albrecht and Norenberg, 2006; Norenberg and Martinez-Hernandez, 1979). Despite this detoxification, an excessive glutamine synthesis is deleterious to astrocytes, as glutamine is transported into and further metabolized in their mitochondria (Albrecht and Norenberg, 2006; Häussinger and Görg, 2010). As a result, the ammonium ions accumulate inside the astrocytes' mitochondria, releasing free radicals and damaging the mitochondrial membrane (Jayakumar et al., 2004; Ziemińska et al., 2000).

One of the main functions of astrocytes is glutamate re-uptake after neural signal transmission. As a consequence, astrocyte malfunction should lead to an extracellular glutamate increase. In line with this, increased glutamate concentrations in the cerebrospinal fluid of patients with cirrhotic liver has already been reported (Monfort et al., 2002; Watanabe et al., 1984), and glutamate excess has been demonstrated in the substantia nigra and the nucleus accumbens of rats with portocaval shunts (Canales et al., 2003; Cauli et al., 2006).

Furthermore, chronic hyperammonemia is related to a change in glutamatergic transduction pathways of both ionotropic and metabotropic glutamate receptors (Cabrera-Pastor et al., 2012; Canales et al., 2003; Hermenegildo et al., 1998). This change of transduction pathways alters glutamatergic neurotransmission in multiple brain regions, which potentially leads to the wide palette of symptoms in HE. For example, hypokinesia and locomotion impairment in HE were linked to changes in the basal ganglia-thalamo-cortical loop (Cauli et al., 2009; Jover et al., 2005), and impaired motor learning in HE had been associated with alteration of the glutamate-NO-cGMP pathway in the cerebellum (Cauli et al., 2007; Erceg et al., 2005). A schematic of the proposed interplay between hyperammonemia and glutamatergic neurotransmission is depicted in **Fig. 5.**

In addition, glutamatergic neurotransmission is closely linked to LTP. In fact, LTP can be interpreted as a chain reaction, in which AMPA- and NMDA-receptor-mediated signal transduction plays a major role (Lee and Kirkwood, 2011; Wen et al., 2013). Here, it seems that ammonia does not directly reduce the postsynaptic AMPA- or NMDA-receptor densities (Palomero-Gallagher et al., 2009; Wen et al., 2013), but rather disturbs the intracellular transduction pathways, on which the two glutamate receptors rely (Llansola et al., 2007; Monfort et al., 2005; Wen et al., 2013). The ammonia-related disturbance of the transduction pathways is probably linked to accumulation of reactive oxygen species (ROS; Albrecht and Norenberg, 2006). The accumulated ROS might target and damage RNA of proteins needed for glutamatergic homeostasis, thus reducing synaptic plasticity (Bemeur et al., 2010; Häussinger and Görg, 2010; Wen et al., 2013). Therefore, the reduced plasticity in HE might be a consequence of disturbances in glutamatergic neurotransmission, which itself is a consequence of ammonia accumulation. However, it is unlikely that hyperammonia and ROS limit their deleterious effects exclusively to glutamatergic homeostasis. As ROS are known to cause RNA oxidation (Schliess et al., 2009), they might also impair multiple neurotransmission systems, including GABA-ergic pathways. Although earlier TMS-studies already reported an altered GABA-ergic tone over M1 and cerebellum in HE (Groiss et al., 2019; Hassan et al., 2019), the exact molecular mechanisms behind those findings are still not entirely understood.

Further, synaptic plasticity and LTP are accepted as one of the leading electrophysiological models for learning and memory. Especially, LTP-like synaptic plasticity in M1 plays a vital role in learning and retention of motor skills (Classen et al., 1998). Indeed, recent studies on mice demonstrated M1 to be the most functionally connected area inside a network of distinct motor areas, which are active when new motor skills are learned (Badea et al., 2019). Importantly, the mechanisms behind motor learning involve temporal plasticity occlusion within M1. In this way, new motor memories are probably formed and protected from interference effects. Importantly, a disrupted synaptic plasticity in M1 is known to decrease motor skill retention in HE patients (Cantarero et al., 2013). It remains up to further investigation to determine the degree of correlation between LTP-decrease and motor learning impairment in patients with manifest HE.

In our study, we also measured CFF, which is largely believed to reflect plasticity changes in the visual system. CFF supports the diagnosis of mHE, where the disease-related symptoms are still too mild to be directly observed by a clinician. Notably, it features 91% sensitivity and 92% specificity in detecting mHE (Metwally et al., 2019) and can also predict Child-Pugh-Class and survival rates of cirrhotic patients (Barone et al., 2018). CFF relies on the measurement of the temporal resolution of visual perception, which is systematically impaired in HE. Under normal conditions (i.e., in healthy individuals), the flickering light is first processed in the primary visual cortex (V1), and after that encoded in the inferior parietal lobule, which probably reflects its conscious perception (Carmel et al., 2006; Hagenbeek et al., 2002). In that context, a reduced temporal resolution of visual perception indicated by a decreased CFF threshold might be associated with impaired cortical synaptic plasticity. Earlier findings revealed that the CFF threshold can be causally reduced by applying inhibitory plasticity TMS protocols over the inferior parietal lobule (Nardella et al., 2014). Moreover, CFF threshold is successfully increased after motion-direction visual training (Seitz et al., 2006). Therefore, the reduced CFF threshold in HE might reflect reduced synaptic plasticity in higher-order visual areas. The link between perceptual impairments captured by the CFF and pathological alterations of neural activity in HE is further strengthened by a positive relationship between CFF threshold and occipital alpha band peak frequency (Baumgarten et al., 2018). In this regard, exploring synaptic plasticity reduction in cortical areas other than M1 in HE is a promising avenue for future research.

Our findings might provide a further puzzle piece in the development of disease-specific motor symptoms. According to the findings of Butz et al. and Timmermann et al., major motor symptoms of HE, such as flapping tremor and ataxia, are probably caused by slowed cortico-muscular and thalamo-cortical oscillations and can manifest even before the development of neuropsychiatric symptoms (Butz et al., 2014, 2010; Rose, 2012; Timmermann et al., 2003, 2002). The diversity of motor symptoms in HE is well matched by the diversity of electrophysiological changes, which is thoroughly debated in the literature (reviewed in Häussinger et al., 2021). Earlier TMS studies suggest an increased RMT and longer central motor transduction time (Nolano et al., 1997), as well as increased GABA-ergic and reduced glutamatergic neurotransmission over M1 in minimal HE (Nardone et al., 2016). Additionally, stage-dependent exploration revealed the GABA-ergic neurotransmission over M1 to be reduced in stage two of manifest HE, but increased in the cerebello-cortical pathways, with both effects correlating with HE severity (Groiss et al., 2019; Hassan et al., 2019). In this context, studying HE with different TMS protocols might provide a valuable link between molecular, electrophysiological, and behavioral changes over the course of the disease. Nevertheless, caution is warranted when attributing neuronal population-level effects to LTP-mechanisms. While LTP-changes represent a microscopic event on the cellular level, PAS25-induced changes represent mechanisms on the system level. Although PAS25-induced changes are similar to LTP-changes and can be predicted through them, both mechanisms are not identical (Fung and Robinson, 2013).

A major challenge of the PAS25 protocol seems to be the large inter-individual variability of the PAS25-effects on motor cortex excitability. In our study, only 44% of the controls responded with increased excitability to PAS25 (i.e., LTP- responders). Notably, the proportion of LTP-responders in the current study is in line with earlier reports, where PAS25 was methodically included (Campana et al., 2019; Müller-Dahlhaus et al., 2008). Therefore, approximately half of the participants, in general, are expected to react with an increased cortical excitability after application of PAS25. Known factors that might influence the inter-individual variability include age (Bhandari et al., 2016), cortical thickness, and gyri orientation (Conde et al., 2012). Recent studies, however, failed to provide factors systematically predicting inter-individual plasticity changes induced by PAS (Minkova et al., 2019). Müller-Dahlhaus et al. (2008) reported a reduced PAS25 response with advanced age, irrespective of its direction (both LTP and LTD-like). Interestingly, in their paper, Müller-Dahlhaus et al. (2008) consider that PAS25 could not only cause LTP, but also LTD effects, depending on the individual participant. Similarly, one third of the participants in our study reacted to PAS25 with LTD. Here, not only LTP, but also LTD could be considered as specific response to PAS25. Both LTP and LTD, together, could therefore be viewed as a general PAS25 response, albeit in opposite directions. Based on our data, however, HE significantly contributed to smaller general PAS25-response, independent of the direction.

In this relation, the neuronal populations influenced by PAS25 are dependent on cerebellar input, and earlier studies have shown that cerebellar stimulation can selectively modulate PAS25-induced plasticity. While cerebellar inhibition increases M1 plasticity, cerebellar excitation is known to decrease it (Popa et al., 2013). Moreover, previous reports have shown an increased GABA-ergic cerebellar tone and decreased cerebellar inhibition over M1 in HE (Hassan et al., 2019). Thus, the decreased cerebellar inhibition in HE might serve as the functional mechanism underlying the reduced synaptic plasticity in M1 we report here. Therefore, it cannot be ruled out that our results are, at least partially, influenced by the choice of protocol. In contrast to PAS25, PAS21.5, and iTBS are known to be cerebellum-independent (Hamada et al., 2012; Popa et al., 2013). The optimal protocol to explore synaptic plasticity in HE is, however, not easy to determine. In this relation, it would be highly relevant to compare variability in synaptic plasticity between different protocols. However, reports of differences in variability, especially between PAS25 and PAS21.5, remain scarce.

Arm length is a factor that influences neural transduction time and therefore potentially interferes with pulse timing in PAS protocols. In the current study, participants' arm length was not recorded. Nevertheless, reference SEP values for the medianus nerve suggest that body height needs to be >190 cm for an afferent pulse transduction time of 25 ms. As none of the study participants exhibited a body height over 190 cm, the present results are within reliable range. Another limitation of our study is that no glutamatergic and GABA-ergic neurotransmitter measurements, such as intracortical facilitation (ICF) and short-interval intracortical inhibition (SICI), could be obtained due to protocol length restrains. Furthermore, we did not record recruitment curves at stimulation intensities below 100% RMT, which would have ensured that the MEP amplitudes below RMT are indeed zero. However, we would like to point out that for all recorded stimulation intensities, we did not find any significant RMT differences between controls and HE-patients.

## 5. Conclusion

Our study shows a reduced long-term plasticity over M1 in patients with HE, including manifest HE, compared to healthy controls of the same age. Further research is needed to determine the relation between altered plasticity and both motor and non-motor symptoms in HE.

**Funding:** This research was funded by "Deutsche Forschungsgemeinschaft" (DFG), as part of the project SFB974 B07. This project has received funding from the European Union's Horizon 2020 research and innovation programme under the Marie Skłodowska-Curie grant agreement No. 795998 (MSCA-IF-GF MSC GF awarded to T.J.B.).

**Conflicts of interest:** PN, TJB, SSH, SNM, NDF, GK, GH, MSJ, MB, AS and SJG report no conflicts of interest to disclose related to this work.

**Data availability:** Data is available from the corresponding author on a reasonable request.

## References

Albrecht J, Norenberg MD. Glutamine: A Trojan horse in ammonia neurotoxicity. Hepatology 2006;44:788–94. https://doi.org/10.1002/hep.21357.

Badea A, Ng KL, Anderson RJ, Zhang J, Miller MI, O'Brien RJ. Magnetic resonance imaging of mouse brain networks plasticity following motor learning. PloS One 2019;14:e0216596. https://doi.org/10.1371/journal.pone.0216596.

Barone M, Shahini E, Iannone A, Viggiani MT, Corvace V, Principi M, et al. Critical flicker frequency test predicts overt hepatic encephalopathy and survival in patients with liver cirrhosis. Dig Liver Dis Off J Ital Soc Gastroenterol Ital Assoc Study Liver 2018;50:496–500. https://doi.org/10.1016/j.dld.2018.01.133.

Baumgarten TJ, Neugebauer J, Oeltzschner G, Füllenbach N-D, Kircheis G, Häussinger D, et al. Connecting occipital alpha band peak frequency, visual temporal resolution, and occipital GABA levels in healthy participants and hepatic encephalopathy patients. NeuroImage Clin 2018;20:347–56. https://doi.org/10.1016/j.nicl.2018.08.013.

Bemeur C, Desjardins P, Butterworth RF. Evidence for oxidative/nitrosative stress in the pathogenesis of hepatic encephalopathy. Metab Brain Dis 2010;25:3–9. https://doi.org/10.1007/s11011-010-9177-y.

Bhandari A, Radhu N, Farzan F, Mulsant BH, Rajji TK, Daskalakis ZJ, et al. A Meta-Analysis of the Effects of Aging on Motor Cortex Neurophysiology Assessed by Transcranial Magnetic Stimulation. Clin Neurophysiol Off J Int Fed Clin Neurophysiol 2016;127:2834–45. https://doi.org/10.1016/j.clinph.2016.05.363.

Butler EG, Horne MK, Rawson JA. Sensory characteristics of monkey thalamic and motor cortex neurones. J Physiol 1992;445:1–24. https://doi.org/10.1113/jphysiol.1992.sp018909.

Butz M, May ES, Häussinger D, Schnitzler A. The slowed brain: Cortical oscillatory activity in hepatic encephalopathy. Arch Biochem Biophys 2013;536:197–203. https://doi.org/10.1016/j.abb.2013.04.004.

Butz M, Timmermann L, Braun M, Groiss SJ, Wojtecki L, Ostrowski S, et al. Motor impairment in liver cirrhosis without and with minimal hepatic encephalopathy. Acta Neurol Scand 2010;122:27–35. https://doi.org/10.1111/j.1600-0404.2009.01246.x.

Butz M, Timmermann L, Gross J, Pollok B, Südmeyer M, Kircheis G, et al. Cortical activation associated with asterixis in manifest hepatic encephalopathy. Acta Neurol Scand 2014;130:260–7. https://doi.org/10.1111/ane.12217.

Cabrera-Pastor A, Llansola M, Reznikov V, Boix J, Felipo V. Differential effects of chronic hyperammonemia on modulation of the glutamate-nitric oxide-cGMP pathway by metabotropic glutamate receptor 5 and low and high affinity AMPA receptors in cerebellum in vivo. Neurochem Int 2012;61:63–71. https://doi.org/10.1016/j.neuint.2012.04.006.

Campana M, Papazova I, Pross B, Hasan A, Strube W. Motor-cortex excitability and response variability following paired-associative stimulation: a proof-of-concept study comparing individualized and fixed inter-stimulus intervals. Exp Brain Res 2019. https://doi.org/10.1007/s00221-019-05542-x.

Canales J-J, Elayadi A, Errami M, Llansola M, Cauli O, Felipo V. Chronic hyperammonemia alters motor and neurochemical responses to activation of group I metabotropic glutamate receptors in the nucleus accumbens in rats in vivo. Neurobiol Dis 2003;14:380–90. https://doi.org/10.1016/j.nbd.2003.08.023.

Cantarero G, Lloyd A, Celnik P. Reversal of long-term potentiation-like plasticity processes after motor learning disrupts skill retention. J Neurosci Off J Soc Neurosci 2013;33:12862–9. https://doi.org/10.1523/JNEUROSCI.1399-13.2013.

Carmel D, Lavie N, Rees G. Conscious awareness of flicker in humans involves frontal and parietal cortex. Curr Biol CB 2006;16:907–11. https://doi.org/10.1016/j.cub.2006.03.055.

Cauli O, Llansola M, Erceg S, Felipo V. Hypolocomotion in rats with chronic liver failure is due to increased glutamate and activation of metabotropic glutamate receptors in substantia nigra. J Hepatol 2006;45:654–61. https://doi.org/10.1016/j.jhep.2006.06.019.

Cauli O, Mlili N, Rodrigo R, Felipo V. Hyperammonaemia alters the mechanisms by which metabotropic glutamate receptors in nucleus accumbens modulate motor function. J Neurochem 2007;103:38–46. https://doi.org/10.1111/j.1471-4159.2007.04734.x.

Cauli O, Rodrigo R, Llansola M, Montoliu C, Monfort P, Piedrafita B, et al. Glutamatergic and gabaergic neurotransmission and neuronal circuits in hepatic encephalopathy. Metab Brain Dis 2009;24:69–80. https://doi.org/10.1007/s11011-008-9115-4.

Classen J, Liepert J, Wise SP, Hallett M, Cohen LG. Rapid plasticity of human cortical movement representation induced by practice. J Neurophysiol 1998;79:1117–23. https://doi.org/10.1152/jn.1998.79.2.1117.

Classen J, Wolters A, Stefan K, Wycislo M, Sandbrink F, Schmidt A, et al. Paired associative stimulation. Suppl Clin Neurophysiol 2004;57:563–9.

Coltart I, Tranah TH, Shawcross DL. Inflammation and hepatic encephalopathy. Arch Biochem Biophys 2013;536:189–96. https://doi.org/10.1016/j.abb.2013.03.016.

Conde V, Vollmann H, Sehm B, Taubert M, Villringer A, Ragert P. Cortical thickness in primary sensorimotor cortex influences the effectiveness of paired associative stimulation. NeuroImage 2012;60:864–70. https://doi.org/10.1016/j.neuroimage.2012.01.052.

Delvendahl I, Jung NH, Kuhnke NG, Ziemann U, Mall V. Plasticity of motor threshold and motor-evoked potential amplitude – A model of intrinsic and synaptic plasticity in human motor cortex? Brain Stimulat 2012;5:586–93. https://doi.org/10.1016/j.brs.2011.11.005.

Desjardins P, Du T, Jiang W, Peng L, Butterworth RF. Pathogenesis of hepatic encephalopathy and brain edema in acute liver failure: Role of glutamine redefined. Neurochem Int 2012;60:690–6. https://doi.org/10.1016/j.neuint.2012.02.001.

Di Lazzaro V, Dileone M, Profice P, Pilato F, Oliviero A, Mazzone P, et al. LTD-like plasticity induced by paired associative stimulation: direct evidence in humans. Exp Brain Res 2009;194:661. https://doi.org/10.1007/s00221-009-1774-9.

Di Lazzaro V, Oliviero A, Pilato F, Saturno E, Dileone M, Mazzone P, et al. The physiological basis of transcranial motor cortex stimulation in conscious humans. Clin Neurophysiol 2004;115:255–66. https://doi.org/10.1016/j.clinph.2003.10.009.

Erceg S, Monfort P, Hernandez-Viadel M, Llansola M, Montoliu C, Felipo V. Restoration of learning ability in hyperammonemic rats by increasing extracellular cGMP in brain. Brain Res 2005;1036:115–21. https://doi.org/10.1016/j.brainres.2004.12.045.

Felipo V. Hepatic encephalopathy: effects of liver failure on brain function. Nat Rev Neurosci 2013;14:851–8. https://doi.org/10.1038/nrn3587.

Ferenci P, Lockwood A, Mullen K, Tarter R, Weissenborn K, Blei AT. Hepatic encephalopathy--definition, nomenclature, diagnosis, and quantification: final report of the working party at the 11th World Congresses of Gastroenterology, Vienna, 1998. Hepatol Baltim Md 2002;35:716–21. https://doi.org/10.1053/jhep.2002.31250.

Fung PK, Robinson PA. Neural field theory of calcium dependent plasticity with applications to transcranial magnetic stimulation. J Theor Biol 2013;324:72–83. https://doi.org/10.1016/j.jtbi.2013.01.013.

Golaszewski S, Langthaler PB, Schwenker K, Florea C, Christova M, Brigo F, et al. Abnormal cortical synaptic plasticity in minimal hepatic encephalopathy. Brain Res Bull 2016;125:200–4. https://doi.org/10.1016/j.brainresbull.2016.07.011.

Götz T, Huonker R, Kranczioch C, Reuken P, Witte OW, Günther A, et al. Impaired evoked and resting-state brain oscillations in patients with liver cirrhosis as revealed by magnetoencephalography. NeuroImage Clin 2013;2:873–82. https://doi.org/10.1016/j.nicl.2013.06.003.

Groiss SJ, Butz M, Baumgarten TJ, Füllenbach N-D, Häussinger D, Schnitzler A. GABA-ergic tone hypothesis in hepatic encephalopathy - Revisited. Clin Neurophysiol Off J Int Fed Clin Neurophysiol 2019;130:911–6. https://doi.org/10.1016/j.clinph.2019.03.011.

Hagenbeek RE, Rombouts SARB, van Dijk BW, Barkhof F. Determination of individual stimulus–response curves in the visual cortex. Hum Brain Mapp 2002;17:244–50. https://doi.org/10.1002/hbm.10067.

Hallett M. Transcranial magnetic stimulation and the human brain. Nature 2000;406:147. https://doi.org/10.1038/35018000.

Hamada M, Strigaro G, Murase N, Sadnicka A, Galea JM, Edwards MJ, et al. Cerebellar modulation of human associative plasticity: Cerebellum and human associative plasticity. J Physiol 2012;590:2365–74. https://doi.org/10.1113/jphysiol.2012.230540.

Hassan SS, Baumgarten TJ, Ali AM, Füllenbach N-D, Jördens MS, Häussinger D, et al. Cerebellar inhibition in hepatic encephalopathy. Clin Neurophysiol Off J Int Fed Clin Neurophysiol 2019;130:886–92. https://doi.org/10.1016/j.clinph.2019.02.020.

Häussinger D, Butz M, Schnitzler A, Görg B. Pathomechanisms in hepatic encephalopathy. Biol Chem 2021. https://doi.org/10.1515/hsz-2021-0168.

Häussinger D, Görg B. Interaction of oxidative stress, astrocyte swelling and cerebral ammonia toxicity. Curr Opin Clin Nutr Metab Care 2010;13:87. https://doi.org/10.1097/MCO.0b013e328333b829.

Häussinger D, Schliess F. Pathogenetic mechanisms of hepatic encephalopathy. Gut 2008;57:1156–65. https://doi.org/10.1136/gut.2007.122176.

Häussinger D, Sies H. Hepatic encephalopathy: clinical aspects and pathogenetic concept. Arch Biochem Biophys 2013;536:97–100. https://doi.org/10.1016/j.abb.2013.04.013.

Henderson JM. Hepatic Encephalopathy. Textb. Hepatol., John Wiley & Sons, Ltd; 2008, p. 728–60. https://doi.org/10.1002/9780470691861.ch7h.

Hermenegildo C, Montoliu C, Llansola M, Muñoz MD, Gaztelu JM, Miñana MD, et al. Chronic hyperammonemia impairs the glutamate-nitric oxide-cyclic GMP pathway in cerebellar neurons in culture and in the rat in vivo. Eur J Neurosci 1998;10:3201–9. https://doi.org/10.1046/j.1460-9568.1998.00329.x.

Huang Y-Z, Edwards MJ, Rounis E, Bhatia KP, Rothwell JC. Theta Burst Stimulation of the Human Motor Cortex. Neuron 2005;45:201–6. https://doi.org/10.1016/j.neuron.2004.12.033.

Jayakumar AR, Rama Rao KV, Schousboe A, Norenberg MD. Glutamine-induced free radical production in cultured astrocytes. Glia 2004;46:296–301. https://doi.org/10.1002/glia.20003.

Jover R, Compañy L, Gutiérrez A, Lorente M, Zapater P, Poveda MJ, et al. Clinical significance of extrapyramidal signs in patients with cirrhosis. J Hepatol 2005;42:659–65. https://doi.org/10.1016/j.jhep.2004.12.030.

Kircheis G, Hilger N, Häussinger D. Value of Critical Flicker Frequency and Psychometric Hepatic Encephalopathy Score in Diagnosis of Low-Grade Hepatic Encephalopathy. Gastroenterology 2014;146:961-969.e11. https://doi.org/10.1053/j.gastro.2013.12.026.

Kircheis G, Wettstein M, Timmermann L, Schnitzler A, Häussinger D. Critical flicker frequency for quantification of low-grade hepatic encephalopathy. Hepatology 2002;35:357–66. https://doi.org/10.1053/jhep.2002.30957.

Lazar M, Butz M, Baumgarten TJ, Füllenbach N-D, Jördens MS, Häussinger D, et al. Impaired Tactile Temporal Discrimination in Patients With Hepatic Encephalopathy. Front Psychol 2018;9:2059. https://doi.org/10.3389/fpsyg.2018.02059.

Lee H-K, Kirkwood A. AMPA receptor regulation during synaptic plasticity in hippocampus and neocortex. Semin Cell Dev Biol 2011;22:514–20. https://doi.org/10.1016/j.semcdb.2011.06.007.

Llansola M, Rodrigo R, Monfort P, Montoliu C, Kosenko E, Cauli O, et al. NMDA receptors in hyperammonemia and hepatic encephalopathy. Metab Brain Dis 2007;22:321–35. https://doi.org/10.1007/s11011-007-9067-0.

López-Alonso V, Cheeran B, Río-Rodríguez D, Fernández-Del-Olmo M. Inter-individual variability in response to non-invasive brain stimulation paradigms. Brain Stimulat 2014;7:372–80. https://doi.org/10.1016/j.brs.2014.02.004.

Malenka RC, Bear MF. LTP and LTD: An Embarrassment of Riches. Neuron 2004;44:5–21. https://doi.org/10.1016/j.neuron.2004.09.012.

Mansvelder HD, Verhoog MB, Goriounova NA. Synaptic plasticity in human cortical circuits: cellular mechanisms of learning and memory in the human brain? Curr Opin Neurobiol 2019;54:186–93. https://doi.org/10.1016/j.conb.2018.06.013.

May ES, Butz M, Kahlbrock N, Brenner M, Hoogenboom N, Kircheis G, et al. Hepatic encephalopathy is associated with slowed and delayed stimulus-associated somatosensory alpha activity. Clin Neurophysiol 2014;125:2427–35. https://doi.org/10.1016/j.clinph.2014.03.018.

Metwally MA, Biomy HA, Omar MZ, Sakr AI. Critical flickering frequency test: a diagnostic tool for minimal hepatic encephalopathy. Eur J Gastroenterol Hepatol 2019;31:1030–4. https://doi.org/10.1097/MEG.0000000000001375.

Minkova L, Peter J, Abdulkadir A, Schumacher LV, Kaller CP, Nissen C, et al. Determinants of Inter-Individual Variability in Corticomotor Excitability Induced by Paired Associative Stimulation. Front Neurosci 2019;13:841. https://doi.org/10.3389/fnins.2019.00841.

Monfort P, Muñoz M-D, ElAyadi A, Kosenko E, Felipo V. Effects of hyperammonemia and liver failure on glutamatergic neurotransmission. Metab Brain Dis 2002;17:237–50. https://doi.org/10.1023/a:1021993431443.

Monfort P, Muñoz M-D, Felipo V. Molecular Mechanisms of the Alterations in NMDA Receptor-Dependent Long-Term Potentiation in Hyperammonemia. Metab Brain Dis 2005;20:265–74. https://doi.org/10.1007/s11011-005-7905-5.

Müller-Dahlhaus JFM, Orekhov Y, Liu Y, Ziemann U. Interindividual variability and age-dependency of motor cortical plasticity induced by paired associative stimulation. Exp Brain Res 2008;187:467–75. https://doi.org/10.1007/s00221-008-1319-7.

Nakamura K, Groiss SJ, Hamada M, Enomoto H, Kadowaki S, Abe M, et al. Variability in Response to Quadripulse Stimulation of the Motor Cortex. Brain Stimulat 2016;9:859–66. https://doi.org/10.1016/j.brs.2016.01.008.

Nakatani-Enomoto S, Hanajima R, Hamada M, Terao Y, Matsumoto H, Shirota Y, et al. Bidirectional modulation of sensory cortical excitability by quadripulse transcranial magnetic stimulation (QPS) in humans. Clin Neurophysiol Off J Int Fed Clin Neurophysiol 2012;123:1415–21. https://doi.org/10.1016/j.clinph.2011.11.037.

Nardella A, Rocchi L, Conte A, Bologna M, Suppa A, Berardelli A. Inferior parietal lobule encodes visual temporal resolution processes contributing to the critical flicker frequency threshold in humans. PloS One 2014;9:e98948. https://doi.org/10.1371/journal.pone.0098948.

Nardone R, De Blasi P, Höller Y, Brigo F, Golaszewski S, Frey VN, et al. Intracortical inhibitory and excitatory circuits in subjects with minimal hepatic encephalopathy: a TMS study. Metab Brain Dis 2016;31:1065–70. https://doi.org/10.1007/s11011-016-9848-4.

Nolano M, Guardascione MA, Amitrano L, Perretti A, Fiorillo F, Ascione A, et al. Cortico-spinal pathways and inhibitory mechanisms in hepatic encephalopathy. Electroencephalogr Clin Neurophysiol 1997;105:72–8. https://doi.org/10.1016/s0924-980x(96)96571-6.

Norenberg MD, Martinez-Hernandez A. Fine structural localization of glutamine synthetase in astrocytes of rat brain. Brain Res 1979;161:303–10. https://doi.org/10.1016/0006-8993(79)90071-4.

Olde Damink SWM, Deutz NEP, Dejong CHC, Soeters PB, Jalan R. Interorgan ammonia metabolism in liver failure. Neurochem Int 2002;41:177–88. https://doi.org/10.1016/S0197-0186(02)00040-2.

Palomero-Gallagher N, Bidmon H-J, Cremer M, Schleicher A, Kircheis G, Reifenberger G, et al. Neurotransmitter Receptor Imbalances in Motor Cortex and Basal Ganglia in Hepatic Encephalopathy. Cell Physiol Biochem 2009;24:291–306. https://doi.org/10.1159/000233254.

Popa T, Velayudhan B, Hubsch C, Pradeep S, Roze E, Vidailhet M, et al. Cerebellar processing of sensory inputs primes motor cortex plasticity. Cereb Cortex 2013;23:305–14.

Rose CF. Ammonia-Lowering Strategies for the Treatment of Hepatic Encephalopathy. Clin Pharmacol Ther 2012;92:321–31. https://doi.org/10.1038/clpt.2012.112.

Rossini PM, Burke D, Chen R, Cohen LG, Daskalakis Z, Di Iorio R, et al. Non-invasive electrical and magnetic stimulation of the brain, spinal cord, roots and peripheral nerves: Basic principles and procedures for routine clinical and research application. An updated report from an I.F.C.N. Committee. Clin Neurophysiol 2015;126:1071–107. https://doi.org/10.1016/j.clinph.2015.02.001.

Rothwell JC. Techniques and mechanisms of action of transcranial stimulation of the human motor cortex. J Neurosci Methods 1997;74:113–22. https://doi.org/10.1016/S0165-0270(97)02242-5.

Sale MV, Ridding MC, Nordstrom MA. Factors influencing the magnitude and reproducibility of corticomotor excitability changes induced by paired associative stimulation. Exp Brain Res 2007;181:615–26. https://doi.org/10.1007/s00221-007-0960-x.

Schliess F, Görg B, Häussinger D. RNA oxidation and zinc in hepatic encephalopathy and hyperammonemia. Metab Brain Dis 2009;24:119. https://doi.org/10.1007/s11011-008-9125-2.

Schroeter A, Wen S, Mölders A, Erlenhardt N, Stein V, Klöcker N. Depletion of the AMPAR reserve pool impairs synaptic plasticity in a model of hepatic encephalopathy. Mol Cell Neurosci 2015;68:331–9. https://doi.org/10.1016/j.mcn.2015.09.001.

Seitz AR, Nanez JE, Holloway SR, Watanabe T. Perceptual learning of motion leads to faster flicker perception. PloS One 2006;1:e28. https://doi.org/10.1371/journal.pone.0000028.

Stefan K, Kunesch E, Benecke R, Cohen LG, Classen J. Mechanisms of enhancement of human motor cortex excitability induced by interventional paired associative stimulation. J Physiol 2002;543:699–708. https://doi.org/10.1113/jphysiol.2002.023317.

Stefan K, Kunesch E, Cohen LG, Benecke R, Classen J. Induction of plasticity in the human motor cortex by paired associative stimulation. Brain 2000;123:572–84. https://doi.org/10.1093/brain/123.3.572.

Stepanova M, Mishra A, Venkatesan C, Younossi ZM. In-Hospital Mortality and Economic Burden Associated With Hepatic Encephalopathy in the United States From 2005 to 2009. Clin Gastroenterol Hepatol 2012;10:1034-1041.e1. https://doi.org/10.1016/j.cgh.2012.05.016.

Strigaro G, Hamada M, Murase N, Cantello R, Rothwell JC. Interaction Between Different Interneuron Networks Involved in Human Associative Plasticity. Brain Stimulat 2014;7:658–64. https://doi.org/10.1016/j.brs.2014.05.010.

Suppa A, Quartarone A, Siebner H, Chen R, Di Lazzaro V, Del Giudice P, et al. The associative brain at work: Evidence from paired associative stimulation studies in humans. Clin Neurophysiol 2017;128:2140–64. https://doi.org/10.1016/j.clinph.2017.08.003.

Tiksnadi A, Murakami T, Wiratman W, Matsumoto H, Ugawa Y. Direct comparison of efficacy of the motor cortical plasticity induction and the interindividual variability between TBS and QPS. Brain Stimulat 2020;13:1824–33. https://doi.org/10.1016/j.brs.2020.10.014.

Timmermann L, Butz M, Gross J, Kircheis G, Häussinger D, Schnitzler A. Neural synchronization in hepatic encephalopathy. Metab Brain Dis 2005;20:337–46. https://doi.org/10.1007/s11011-005-7916-2.

Timmermann L, Butz M, Gross J, Ploner M, Südmeyer M, Kircheis G, et al. Impaired cerebral oscillatory processing in hepatic encephalopathy. Clin Neurophysiol 2008;119:265–72. https://doi.org/10.1016/j.clinph.2007.09.138.

Timmermann L, Gross J, Butz M, Kircheis G, Häussinger D, Schnitzler A. Mini-asterixis in hepatic encephalopathy induced by pathologic thalamo-motor-cortical coupling. Neurology 2003;61:689–92. https://doi.org/10.1212/01.wnl.0000078816.05164.b1.

Timmermann L, Gross J, Kircheis G, Häussinger D, Schnitzler A. Cortical origin of mini-asterixis in hepatic encephalopathy. Neurology 2002;58:295. https://doi.org/10.1212/WNL.58.2.295.

Vilstrup H, Amodio P, Bajaj J, Cordoba J, Ferenci P, Mullen KD, et al. Hepatic encephalopathy in chronic liver disease: 2014 Practice Guideline by the American Association for the Study Of Liver Diseases and the European Association for the Study of the Liver. Hepatology 2014;60:715–35. https://doi.org/10.1002/hep.27210.

Watanabe A, Takei N, Higashi T, Shiota T, Nakatsukasa H, Fujiwara M, et al. Glutamic acid and glutamine levels in serum and cerebrospinal fluid in hepatic encephalopathy. Biochem Med 1984;32:225–31. https://doi.org/10.1016/0006-2944(84)90076-0.

Wen S, Schroeter A, Klöcker N. Synaptic plasticity in hepatic encephalopathy – A molecular perspective. Arch Biochem Biophys 2013;536:183–8. https://doi.org/10.1016/j.abb.2013.04.008.

Wiethoff S, Hamada M, Rothwell JC. Variability in response to transcranial direct current stimulation of the motor cortex. Brain Stimulat 2014;7:468–75. https://doi.org/10.1016/j.brs.2014.02.003.

Wischnewski M, Schutter DJLG. Efficacy and time course of paired associative stimulation in cortical plasticity: Implications for neuropsychiatry. Clin Neurophysiol Off J Int Fed Clin Neurophysiol 2016;127:732–9. https://doi.org/10.1016/j.clinph.2015.04.072.

Wolters A, Sandbrink F, Schlottmann A, Kunesch E, Stefan K, Cohen LG, et al. A temporally asymmetric Hebbian rule governing plasticity in the human motor cortex. J Neurophysiol 2003;89:2339–45. https://doi.org/10.1152/jn.00900.2002.

World Medical Association. World Medical Association Declaration of Helsinki: ethical principles for medical research involving human subjects. JAMA 2013;310:2191–4. https://doi.org/10.1001/jama.2013.281053.

Ziemińska E, Dolińska M, Lazarewicz JW, Albrecht J. Induction of permeability transition and swelling of rat brain mitochondria by glutamine. Neurotoxicology 2000;21:295–300.

**Table 1:** Demographic data, CFF, and MEP ratio in each session (5min, 15min, and 25min post PAS25, respectively), shown in groups sorted after response. The participants were divided in four groups according to their PAS25 response: LTP-responders, LTD-responders; LTP+LTD responders, and all participants. Values represent mean ± SEM. **Abbreviations:** CFF-critical flicker frequency, MEP-motor evoked potential, PAS25-paired associative stimulation with an inter-stimulus interval of 25 ms, LTP-long-term potentiation, LTD-long-term depression.

| | **Controls n / %** | **HE n / %** | **Age (years) Controls** | **Age (years) HE** | **CFF (Hz) Controls** | **CFF (Hz) HE** | **MEP-ratio 5/15/25 m in Controls** | **MEP-ratio 5/15/25 m in HE** |
|---|---|---|---|---|---|---|---|---|
| **PAS25$_{LTP}$-response** | 10 / 44% | 11 / 48% | 60 ± 2.53 | 60.82 ± 1.5 | 41.23 ± 1.14 | 37.2 ± 0.8 | 2.34 ± 0.26 | 1.13 ± 0.14 |
| | | | | | | | 1.74 ± 0.19 | 1.35 ± 0.18 |
| | | | | | | | 1.76 ± 0.24 | 1.35 ± 0.13 |
| **PAS25$_{LTD}$-response** | 10 / 44% | 7 / 30% | 63 ±2.02 | 60.86 ± 2.96 | 42.16 ±0.84 | 36.33 ± 1.21 | 0.59 ± 0.09 | 0.85 ± 0.19 |
| | | | | | | | 0.57 ± 0.05 | 0.68 ± 0.16 |
| | | | | | | | 0.68 ± 0.09 | 0.71 ± 0.09 |
| | 20 / 87% | 18 / 78% | 61.58 ± 1.59 | 60.83 ± 1.42 | 41.74 ± 0.66 | 36.86 ± 0.66 | 1.46 ± 0.24 | 1.02 ± 0.12 |
| **PAS25$_{LTPorLTD}$-response** | | | | | | | 1.16 ± 0.16 | 1.09 ± 0.15 |
| | | | | | | | 1.22 ± 0.18 | 1.1 ± 0.14 |
| **Responders and Non-responders** | 23 / 100% | 23 / 100 % | 61.45±1.46 | 60.83±1.35 | 42.06±0.58 | 36.72±0.54 | 1.41±0.21 | 0.98±0.09 |
| | | | | | | | 1.14±0.14 | 1.07±0.12 |
| | | | | | | | 1.18±0.16 | 1.1±0.09 |

**Figures:**

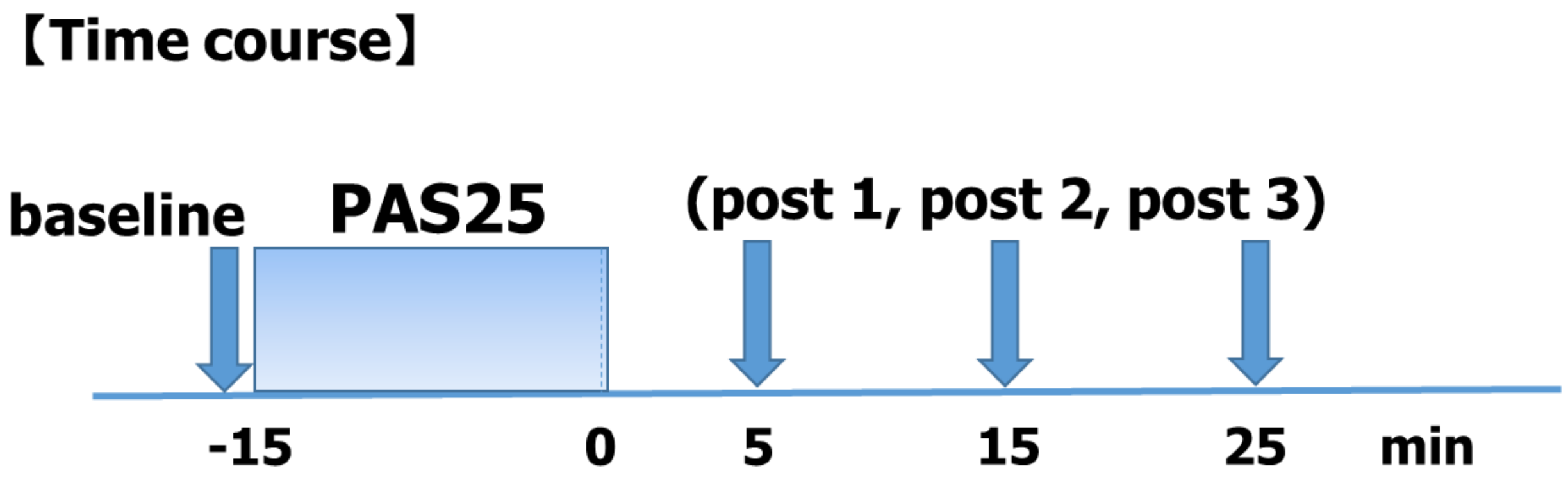

**Figure 1:** Study design: the PAS25 intervention, which lasted 15 minutes, was performed after the baseline session. The next session was carried out 5 minutes after the end of PAS (post1). Two further sessions were carried out in intervals of ten minutes (post2 and post3).

**Abbreviations:** PAS25-paired associative stimulation with an inter-stimulus interval of 25 ms.

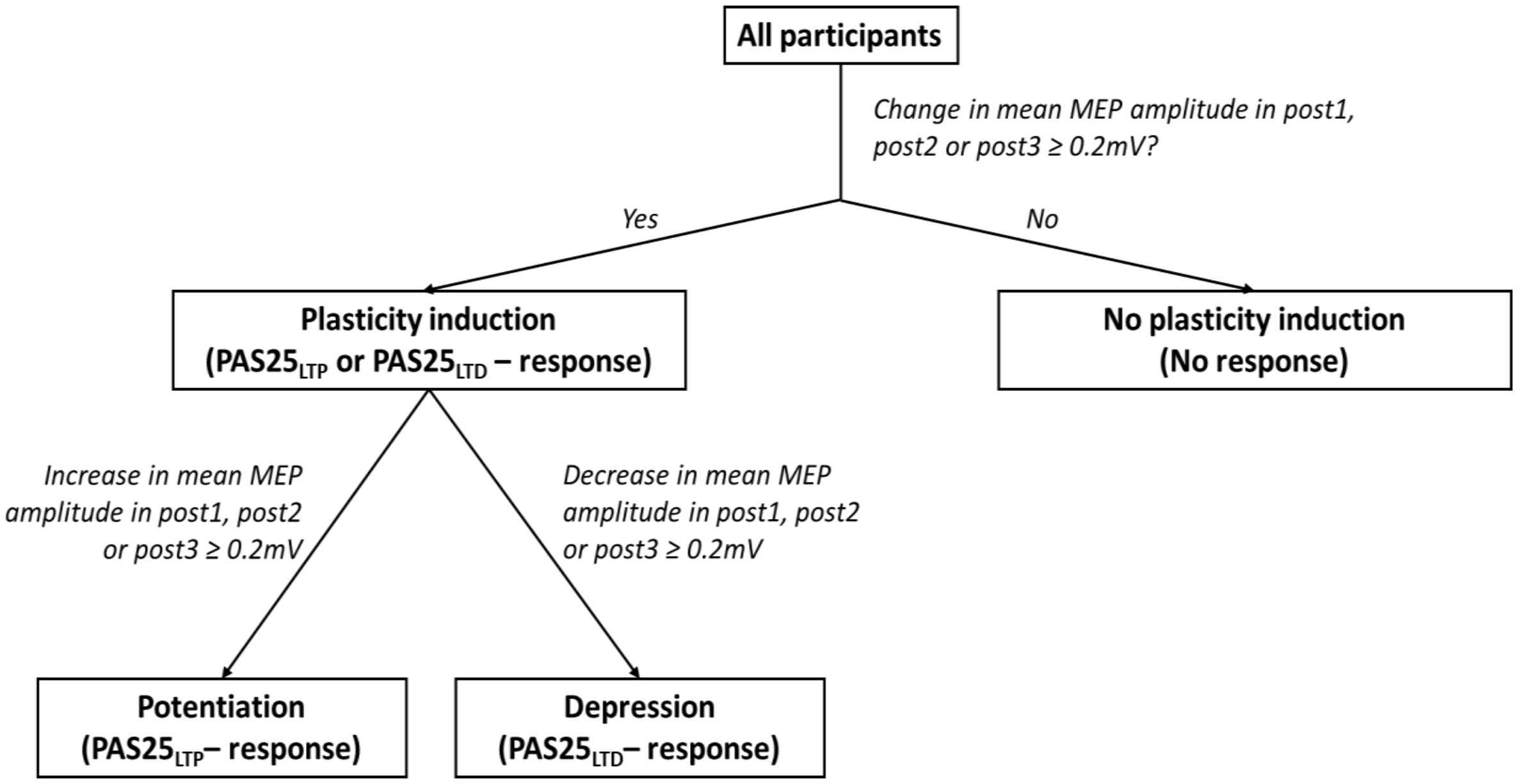

**Figure 2:** Schematics of the subgroup categorization, sorted by response. If a mean MEP change of ≥0.2mV was observed in any of the three post sessions, plasticity induction was assumed to take place. If the absolute maximum MEP amplitude increased by ≥0.2mV compared to baseline, it was considered as LTP. If the absolute maximum MEP amplitude was decreased by ≥0.2mV compared to baseline, it was considered as LTD.

**Abbreviations:** MEP-motor evoked potential, PAS-paired associative stimulation, LTP-long-term potentiation, LTD-long-term depression.

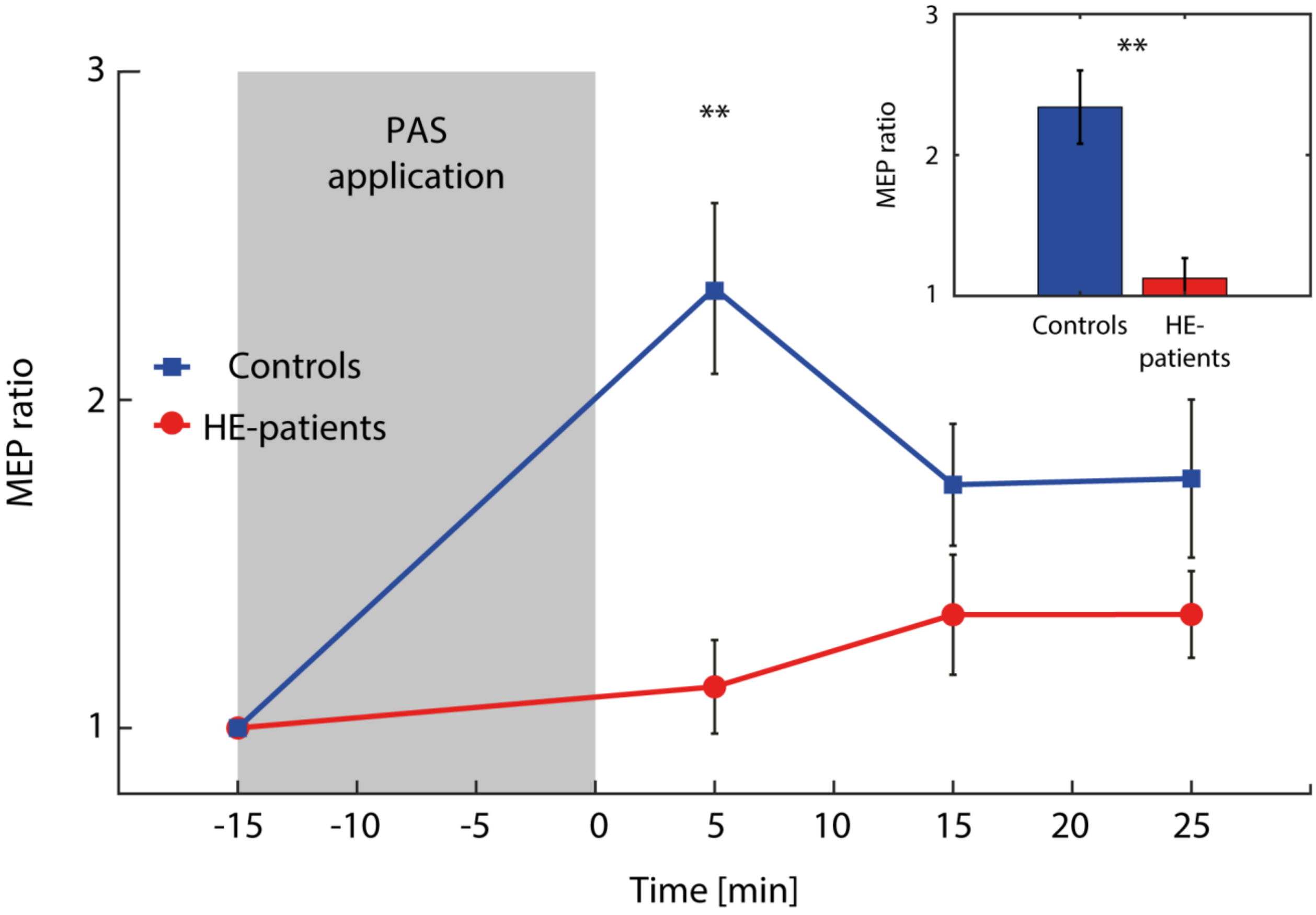

**Figure 3:** MEP ratio for HE-patients and healthy controls, compared to baseline (-15min) in each of the three post PAS25 sessions. MEP-amplitudes at 5 min post PAS25 were significantly potentiated compared to baseline in healthy controls (** p<0.01). Inset: Comparison of MEP ratio at 5 min post PAS25 between HE patients and healthy controls. MEP ratio in the control group was significantly higher than in HE (** p<0.01).

Only LTP responders were included in the analysis. Error bars represent SEM. **Abbreviations:** HE-hepatic encephalopathy, MEP-motor evoked potential, PAS25- paired associative stimulation with an inter-stimulus interval of 25 ms.

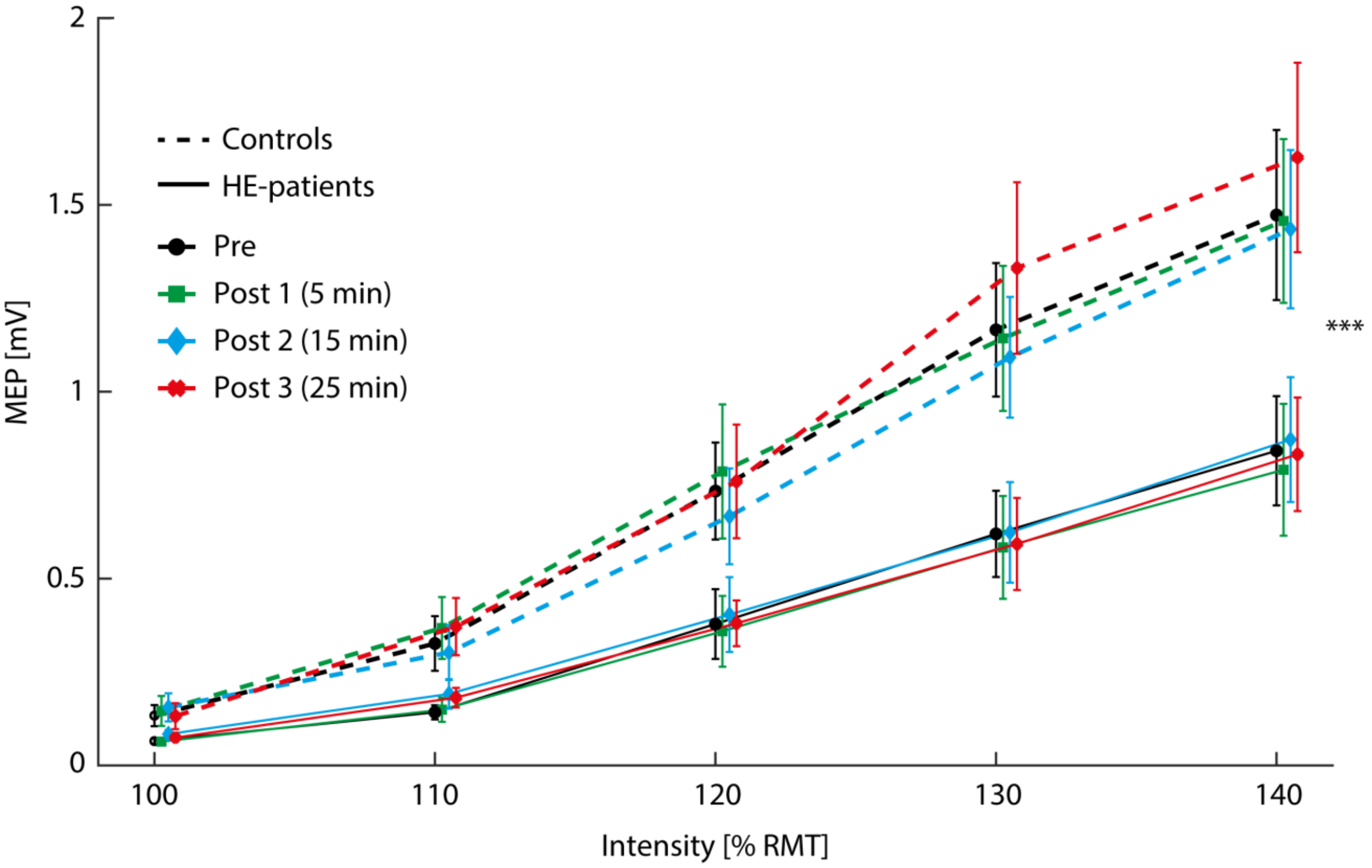

**Figure 4:** Recruitment curves in HE patients and healthy controls across different sessions. MEP amplitudes significantly differ between HE patients and controls (*** p<0.001). In the recruitment curves analysis between the two groups, MEP amplitudes were taken from all intensities and all sessions.

All participants were included in the analysis. Error bars indicate SEM. **Abbreviations:** HE-hepatic encephalopathy, RMT-resting motor threshold, Pre = baseline session before PAS25, post1 = session 5min after PAS25, post2 = session 15min after PAS25, post3 = session 25min after PAS25.

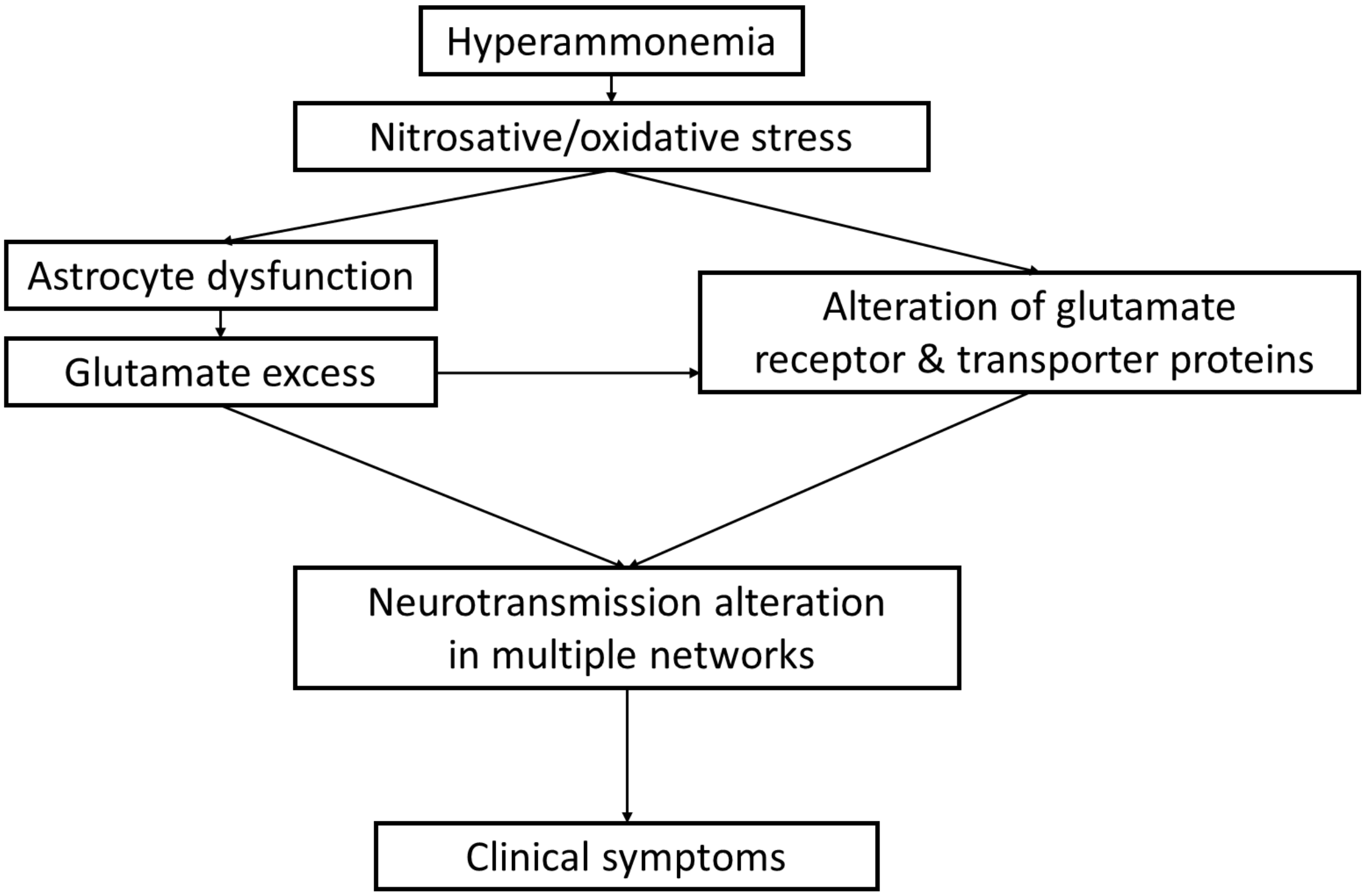

**Figure 5:** Schematics of the relationship between hyperammonemia and glutamatergic neurotransmission, and their putative role in the pathophysiology of Hepatic Encephalopathy.